\documentclass[prl,aps,twocolumn,showpacs,amsmath,amssymb]{revtex4}

\usepackage{graphicx}
\usepackage{dcolumn}
\usepackage{bm}

\def\eep{$(e,e^\prime p)$}
\def\evepv{$(\vec e,e^\prime \vec p\,)$}
\def\evenv{$(\vec e,e^\prime \vec n\,)$}

\def\hevepv{$p(\vec e,e^\prime \vec p\,)$}
\def\devepv{$d(\vec e,e^\prime \vec p\,)n$}
\def\dveven{$\vec d(\vec e,e^\prime n)p$}
\def\devenv{$d(\vec e,e^\prime \vec n\,)p$}

\def\bolddevepv{${\bm d(\vec e,e^\prime \vec p\,)}n$}

\def\gevsq{(GeV/c)$^2$}
\def\px{$P_x^\prime$}
\def\pz{$P_z^\prime$}
\def\polrat{$P_x^\prime/P_z^\prime$}

\def\doublerat{$(P_x^\prime/P_z^\prime)_D/(P_x^\prime/P_z^\prime)_H$}
\def\fdoublerat{$\frac{(P_x^\prime/P_z^\prime)_D}{(P_x^\prime/P_z^\prime)_H}$}
\def\comma{,\ }

\begin{document}

\title{
Polarization transfer in the \bolddevepv{} reaction up to $Q^2$=1.61 \gevsq}

\author {
B.~Hu,$^{1}$
M.K.~Jones,$^{2}$
P.E.~Ulmer,$^{2}$
H.~Arenh\"ovel,$^{3}$
O.K.~Baker,$^{1}$
W.~Bertozzi,$^{4}$
E.J.~Brash,$^{5}$
J.~Calarco,$^{6}$
J.-P.~Chen,$^{7}$
E.~Chudakov,$^{7}$
A.~Cochran,$^{1}$
S.~Dumalski,$^{5}$
R.~Ent,$^{7,1}$
J.M.~Finn,$^{8}$
F.~Garibaldi,$^{9}$
S.~Gilad,$^{4}$
R.~Gilman,$^{10,7}$
C.~Glashausser,$^{10}$
J.~Gomez,$^{7}$
V.~Gorbenko,$^{11}$
J.-O.~Hansen,$^{7}$
J.~Hovebo,$^{5}$
C.W.~de~Jager,$^{7}$
S.~Jeschonnek,$^{12}$
X.~Jiang,$^{10}$
C.~Keppel,$^{1}$
A.~Klein,$^{2}$
A.~Kozlov,$^{5}$
S.~Kuhn,$^{2}$
G.~Kumbartzki,$^{10}$
M.~Kuss,$^{7}$
J.J.~LeRose,$^{7}$
M.~Liang,$^{7}$
N.~Liyanage,$^{4}$
G.J.~Lolos,$^{5}$
P.E.C.~Markowitz,$^{13}$
D.~Meekins,$^{14}$
R.~Michaels,$^{7}$
J.~Mitchell,$^{7}$
Z.~Papandreou,$^{15}$
C.F.~Perdrisat,$^{8}$
V.~Punjabi,$^{16}$
R.~Roche,$^{14}$
D.~Rowntree,$^{4}$
A.~Saha,$^{7}$
S.~Strauch,$^{10}$
L.~Todor,$^{2}$
G.~Urciuoli,$^{9}$
L.B.~Weinstein,$^{2}$
K.~Wijesooriya,$^{8}$
B.B.~Wojtsekhowski,$^{7}$
R.~Woo$^{17}$
\vspace{0.1in}
}

\affiliation{
$^{1}$Hampton University\comma Hampton\comma VA 23668\\
$^{2}$Old Dominion University\comma Norfolk\comma VA 23529\\
$^{3}$Johannes Gutenberg-Universit\"at\comma D-55099 Mainz\comma
Germany\\
$^{4}$Massachusetts Institute of Technology\comma Cambridge\comma MA
02139\\
$^{5}$University of Regina\comma Regina\comma SK\comma Canada S4S 0A2 \\ 
$^{6}$University of New Hampshire\comma Durham\comma NH 03824\\
$^{7}$Thomas Jefferson National Accelerator Facility\comma Newport
News\comma VA 23606\\
$^{8}$College of William and Mary\comma Williamsburg\comma VA 23187\\
$^{9}$INFN\comma Sezione Sanit\'a and Istituto Superiore di Sanit\'a\comma
Laboratorio di Fisica\comma I-00161 Rome\comma Italy\\
$^{10}$Rutgers\comma The State University of New Jersey\comma
Piscataway\comma NJ 08855\\
$^{11}$Kharkov Institute of Physics and Technology\comma Kharkov 310108\comma
Ukraine\\
$^{12}$The Ohio State University\comma Lima\comma OH 45804\\
$^{13}$Florida International University\comma Miami\comma FL 33199\\
$^{14}$Florida State University\comma Tallahassee\comma FL 32306\\
$^{15}$The George Washington University\comma Washington\comma DC
20052\\
$^{16}$Norfolk State University\comma Norfolk\comma VA 23504\\
$^{17}$TRIUMF\comma Vancouver\comma B.C.\comma Canada V6T 2A3\\ 
}

\date{\today}


\begin{abstract}

The recoil proton polarization was measured in the \devepv{} reaction in Hall A
of the  Thomas Jefferson National  Accelerator Facility (JLab).   The electron
kinematics  were centered  on the  quasielastic peak  ($x_{Bj}\approx  1$) and
included three values of  the squared four-momentum transfer, $Q^2$=0.43, 1.00
and 1.61  \gevsq{}.  For $Q^2$=0.43  and 1.61 \gevsq{}, the  missing momentum,
$p_m$, was centered at zero while  for $Q^2$=1.00 \gevsq{} two values of $p_m$
were  chosen:  0   and  174  MeV/c.   At low $p_m$, the $Q^2$ dependence of the
longitudinal polarization, \pz{}, is not well described by a state-of-the-art
calculation.   Further, at higher $p_m$, a 3.5$\sigma$ discrepancy 
was observed in the transverse polarization, \px{}.
Understanding the origin of these discrepancies is important 
in order to confidently extract the neutron electric form
factor from the analogous  \devenv{} experiment.

\end{abstract}

\pacs{25.30.Fj, 13.40.Gp, 13.88.+e, 14.20.Dh} 

%


\maketitle

In the loosely  bound deuteron, the proton and neutron  are expected to behave
essentially as  free particles in  intermediate energy nuclear  reactions with
appropriate  kinematics.  This expectation  and the  absence of  suitable pure
neutron targets make  the deuteron a natural choice  for extracting properties
of  the neutron.  Though  the neutron  elastic electric  form factor  has been
especially difficult  to extract,  the use of  polarized beams and  targets in
\dveven{} \cite{Pa99,Wa04}  and polarized beams with neutron  
recoil polarimetry in
\devenv{} \cite{Ed94,He99,Os99,Ma03,Gl05} has allowed statistically precise 
measurements. 

For  elastic  electron  scattering  from  a  free nucleon,  it  was  shown  in
\cite{AR74,ACG81} that the polarizations transferred from a 
longitudinally polarized
electron  beam to  the  recoil nucleon  (i.e.,  via the  \evepv{} or  \evenv{}
reaction)  can be  expressed  in  terms of  the  nucleon electromagnetic  form
factors.  This technique has been  exploited to measure the proton electric to
magnetic  form factor  ratio for  large  values of  the squared  four-momentum
transfer, $Q^2$,  using a hydrogen target \cite{Jo00,Ga02,Pu05}.   In order to
extract  the neutron  electric form  factor, the  \devenv{} reaction  has been
exploited  at  the MIT-Bates  Laboratory  \cite{Ed94},  
Mainz \cite{He99,Os99,Gl05}  and
Jefferson Lab (JLab) \cite{Ma03}.  However, nuclear effects can compromise the
direct  connection  between the  polarization  transfer  coefficients and  the
neutron  form factors.   This is especially true of the neutron electric form
factor, given its small size relative to possible competing effects.
It is  therefore  essential that  reaction models  be
tested experimentally.   The  present  experiment,  employing  the  \devepv{}
reaction,  provides the means  for evaluating the validity of extracting form
factors from the polarization transfer coefficients, since  the polarization
observables can  be compared directly with  those obtained from  a free proton
target via the elastic \hevepv{}  reaction.  (In addition, our data may provide
useful   information  for   the   related  $^4$He\evepv{}$^3$H experiments
\cite{Di01,e93049,e03104},
where the higher nuclear  density likely leads to
more important nuclear effects.)

\begin{figure}
\includegraphics[width=30pc]{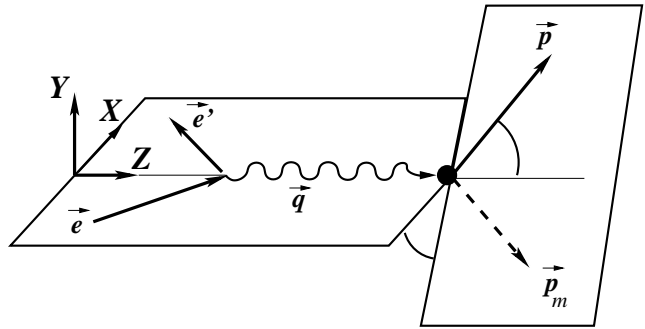}
\caption{\label{fig:kin} Coordinate system used to define the
polarization components.  The $Z$ axis is along the momentum transfer
$\vec q$, the $Y$ axis is in the direction of $\vec e\,\times\vec
e\,^\prime$ (where $\vec e\,$ and $\vec e\,^\prime$ are the momenta of
the incident and scattered electron, respectively) and the $X$ axis is
in the electron scattering plane completing the right-handed system.
Here, $\vec p\,$ is the momentum of the recoiling proton, and $\vec
p_m\,$ is the missing momentum.  The ``out-of-plane'' angle is the angle 
between the two depicted planes, the scattering plane and the hadronic plane.}
\end{figure}

In  the simplest  picture of  the \devepv{}  reaction, the  plane wave
impulse approximation (PWIA), the proton is knocked out by the virtual
photon  and  is detected  without  any  further  interaction with  the
unobserved  neutron.  In this  picture, the  transferred polarizations
(see Fig.~\ref{fig:kin} for an illustration of the coordinate system)
along the  momentum transfer direction, \pz{}, and in  the scattering
plane, perpendicular to  the momentum transfer, \px{},
can be expressed in terms  of various kinematical factors and the ratio of the
proton electric  and magnetic  form  factors ($G_E$  and $G_M$,  respectively)
\cite{PVo89}. Various calculations \cite{FA79,Ar87,Ko88,Re89,La90,Mo91,Arpriv}
predict  that
polarizations measured  in the  \devenv{} and \devepv{}  reactions for
kinematics  close  to  zero   missing  momentum  ($p_m$,  where  $\vec
p_m\equiv \vec  q-\vec p\,$ with $\vec q$  the three-momentum transfer
and $\vec p$  the momentum of the detected nucleon)  are expected to be
nearly free  from the effects of interaction  currents [meson exchange
currents (MEC) and isobar configurations  (IC)] as well as final-state
interactions (FSI) between the outgoing nucleons.  It is precisely the
predicted  insensitivity  to such  effects  which  made the  \devenv{}
reaction a  natural choice for  the extraction of the  neutron electric
form factor.  However, the moderate experimental acceptances employed
in these  experiments entail an average  over kinematics outside
the  ideal  limit of  $p_m=0$.   Polarizations
measured  in  the  \devepv{}  reaction  can test  some  of  the  model
assumptions over the kinematical range of interest.

To date  only two  other experiments on  the \devepv{}  reaction exist,
one  performed at the Mainz Microtron
(MAMI) facility \cite{Ey95} and the  other at the MIT-Bates Laboratory 
\cite{Mi98}.  They were 
restricted to squared
four-momentum transfers of $Q^2$=0.3 \gevsq{} (Mainz) and
$Q^2$=0.38  and 0.50 \gevsq{} (Bates) and also to low $p_m$.
The data from both experiments were well  described by theoretical models.
The  current JLab  experiment was  able to  achieve
higher $Q^2$ and $p_m$ values with smaller statistical uncertainties.

Three of  our kinematics settings  were centered at $p_m=0$,  roughly covering
the  $Q^2$ range of  the JLab  \devenv{} experiment  \cite{Ma03}.  At  each of
these  kinematics, both  \devepv{}  and \hevepv{}  data  were acquired.   This
allowed  forming  ratios  of  the  polarizations for  deuterium  and  hydrogen
targets,  providing a  measure of  nuclear effects.   A fourth  kinematics was
selected at non-zero $p_m$, at the  intermediate $Q^2$ value, in order to test
reaction  models in  a region  where interaction  effects are  expected  to be
somewhat  larger.  Furthermore, this kinematics  is relevant for the \devenv{}
experiment given  that its acceptance includes $p_m$ values of this magnitude.

The  experiment  was  performed in  Hall  A  of  JLab using  the  high
resolution  spectrometer pair.   The relevant kinematical parameters are 
given  in Table~\ref{kin_tbl}. Details  of the Hall A instrumentation
are given elsewhere \cite{nim}.  Electrons were detected in the ``Left''
spectrometer while protons were detected in the ``Right'' spectrometer.
The targets consisted of 15 cm long liquid hydrogen and deuterium cells.
The Left spectrometer included an atmospheric pressure CO$_2$ {\v C}erenkov
detector used to reject $\pi^-$ events.
In order to reduce other backgrounds, nominal cuts were placed 
on the vertex and angular variables reconstructed at the target.
Uncorrelated $ep$ coincidences were removed via cuts on the coincidence
time-of-flight, as well as cuts on the missing mass and missing
momentum.
The  experiment used beam  currents of up  to 50
$\mu$A combined  with a beam  polarization of 76\%, measured
using  a  M{\o}ller  polarimeter.    The  beam  helicity  was  flipped
pseudo-randomly  to reduce  systematic uncertainties  of  the extracted
polarization  transfer  observables.    The  proton  spectrometer  was
equipped with  a focal plane  polarimeter (FPP) \cite{Pu05}.
Polarized  protons scatter  azimuthally asymmetrically  in  the carbon
analyzer of the FPP.  The analyzer thicknesses employed are given in
Table~\ref{fpp_tbl}.  In order to reduce Coulomb scattering for which the
analyzing power is identically zero, cuts restricting the polar angle of the
second-scattering distribution were enforced and are shown in
Table~\ref{fpp_tbl}.  
The resulting distributions, in combination with
information on the beam helicity,  were analyzed by means of a maximum
likelihood method  to obtain the transferred polarization
components.  
More details  on  the  analysis can  be  found in  
Refs. \cite{Pu05,sonjaphd}.

\begin{table}
\caption{\label{kin_tbl}  Kinematics (central values) for the present 
experiment.  The beam energy was 1.669 GeV for all kinematics.}
\begin{ruledtabular}
\begin{tabular}{cccccc}
 $Q^2$ & $p_m$ & Electron & Electron       & Proton   & Proton        \\
       &       & Momentum & $\theta_{LAB}$ & Momentum &$\theta_{LAB}$ \\
 \gevsq{} & MeV/c & GeV/c & degrees & GeV/c  & degrees    \\
\hline
 0.43   &   0 & 1.429    & 24.45     & 0.692    & $-$58.97 \\
 1.00   &   0 & 1.127    & 42.65     & 1.128    & $-$42.68 \\
 1.61   &   0 & 0.804    & 66.23     & 1.525    & $-$28.91 \\
 1.00   & 174 & 1.127    & 42.65     & 1.128    & $-$33.88 \\
\end{tabular}
\end{ruledtabular}
\end{table}

\begin{table}
\caption{\label{fpp_tbl}  Thickness of the FPP graphite analyzer for each of
our kinematics.  Also shown are the cuts we placed on the polar angle of the
second scattering in the FPP.
}
\begin{ruledtabular}
\begin{tabular}{cccccc}
 $Q^2$ & $p_m$    & Analyzer Thickness & $\theta_{\rm FPP}$ Cut \\
 \gevsq{} & MeV/c & inches             &  degrees               \\
\hline
 0.43   &   0 & 3.0  & 3--30 \\
 1.00   &   0 & 9.0  & 3--30 \\
 1.61   &   0 & 16.5 & 3--40 \\
 1.00   & 174 & 9.0  & 3--30 \\
\end{tabular}
\end{ruledtabular}
\end{table}

\begin{figure}
\vskip 0.25in
\includegraphics[width=18pc,angle=0]{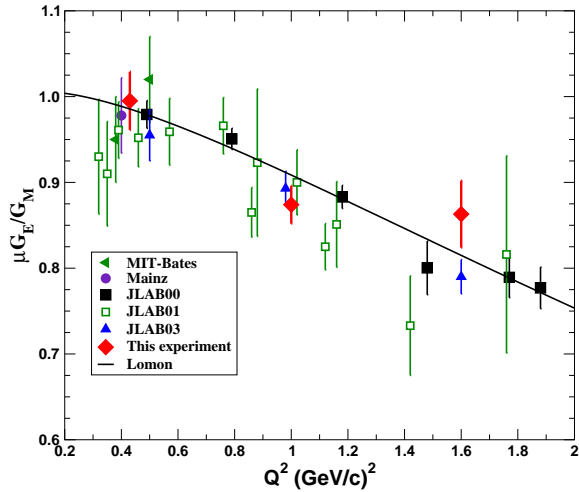}
\caption{\label{fig:hrat} (Color online)
The filled diamonds are $\mu G_E/G_M$  for this experiment.  Data from other 
Jefferson Lab experiments are labeled as JLAB00~\cite{Pu05}, JLAB01~\cite{Ga01}
and JLAB03~\cite{sonjaphd}.  Data from other laboratories are labeled as 
MIT-Bates~\cite{Mi98} and Mainz~\cite{Di01}.  The curve shows $\mu G_E/G_M$
for the Lomon GKex(02S) form factors~\cite{lomon}.}
\end{figure}

As  a check, our  \hevepv{} data  were compared  with the  extracted $G_E/G_M$
ratio  from  previous experiments  which  also  used  the recoil  polarization
technique.   Our  results, listed  in  Table~\ref{ffrat_tbl}  and plotted  as
filled diamonds in  Fig.~\ref{fig:hrat}, are seen to agree  well with previous
measurements.  Also  shown in Fig.~\ref{fig:hrat}  is $\mu G_E/G_M$  for the
Lomon GKex(02S) form factors \cite{lomon}.   The Lomon form factors agree well
with the  polarization transfer  data in this  $Q^2$ range and  were therefore
incorporated in our \devepv{} calculations (see below).

\begin{table}
\caption{\label{ffrat_tbl}
The form factor ratio obtained from our \hevepv{} data, scaled by the proton 
magnetic moment, $\mu$.  The uncertainties are statistical and systematic 
respectively.}
\begin{ruledtabular}
\begin{tabular}{cc}
  $Q^2$   & $\mu G_E/G_M$  \\
 \gevsq{} &                \\ \hline
\hline
 0.43     & 0.994  $\pm$ 0.034 $\pm$ 0.005  \\
 1.00     & 0.879  $\pm$ 0.022 $\pm$ 0.013  \\
 1.61     & 0.865  $\pm$ 0.039 $\pm$ 0.036  \\
\end{tabular}
\end{ruledtabular}
\end{table}

\begin{figure}
\vskip 0.25in
\includegraphics[width=18pc,angle=0]{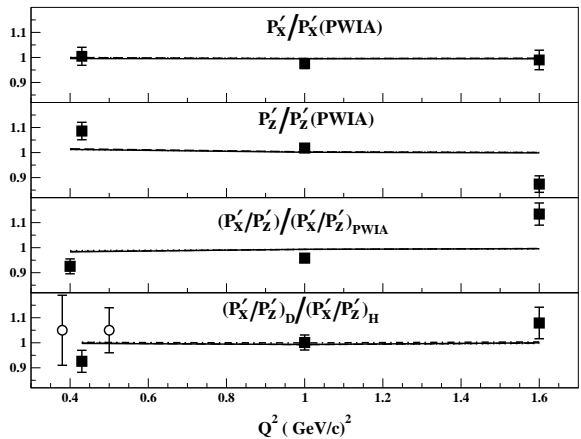}
\caption{\label{fig:ptpl}  The open circles are  the
MIT-Bates data \cite{Mi98} and the filled squares represent the data from
the present experiment.  The dot-dashed curves
are for PWBA, the dotted curves are for DWBA, the dashed curves
include MEC and IC and the solid curves are the full calculations which also
include relativistic corrections (RC).   
The top two panels show \px{} and \pz{}, normalized to the PWIA calculation.
The third panel shows \polrat{} compared to the same ratio calculated in PWIA.
The bottom panel shows the double ratio, defined in the text.}
\end{figure}

\begin{table}
\caption{\label{polrat_tbl} The polarizations, \px{} and \pz{}, and the ratio,
\polrat{}, as a function of $Q^2$ for the \devepv{} measurements centered 
at $p_m=0$.  Also shown are the double ratios, defined in the text. 
The uncertainties are statistical and systematic respectively.  
For \px{} and \pz{}, the statistical uncertainty includes a contribution 
from the statistical uncertainty in our extraction of the analyzing power, 
$A_c$, amounting to $\Delta A_c/A_c=2.7$\%, 1.4\% and 2.3\% for $Q^2=0.43$, 
1.00 and 1.61 respectively.}
\begin{ruledtabular}
\begin{tabular}{ccc}
 $Q^2$    & \px{}                          & \pz{}                          \\
 \gevsq{} &               &               \\ \hline
 0.43   & $-$0.218 $\pm$ 0.008 $\pm$ 0.0006 & 0.236 $\pm$ 0.008 $\pm$ 0.0009 \\
 1.00   & $-$0.299 $\pm$ 0.006 $\pm$ 0.003  & 0.557 $\pm$ 0.009 $\pm$ 0.003 \\
 1.61   & $-$0.279 $\pm$ 0.011 $\pm$ 0.011  & 0.722 $\pm$ 0.024 $\pm$ 0.004 \\
\hline \\
        & \polrat{}                         & \doublerat{}              \\
\hline
 0.43   & $-$0.924 $\pm$ 0.029 $\pm$ 0.005  & 0.926 $\pm$ 0.044 $\pm$ 0.0005 \\
 1.00   & $-$0.537 $\pm$ 0.010 $\pm$ 0.008  & 1.001 $\pm$ 0.030 $\pm$ 0.0007 \\
 1.61   & $-$0.387 $\pm$ 0.015 $\pm$ 0.016  & 1.077 $\pm$ 0.070 $\pm$ 0.0015 \\
\end{tabular}
\end{ruledtabular}
\end{table}

Fig.~\ref{fig:ptpl} and  Table~\ref{polrat_tbl} show results for  
the  three  measurements centered at  $p_m=0$.
The top three panels show \px{}, \pz{} and \polrat{} compared to the PWIA
calculation.  The bottom panel shows the double ratio, \doublerat{}, defined 
as  the ratio \px/\pz{}  for \devepv{} divided  by the same  ratio for
\hevepv{}.  Only  statistical uncertainties  are shown  in the
figure; the systematic uncertainties are  given in the table and are discussed
in detail later in the paper.
The calculations shown are  from Arenh\"ovel \cite{Arpriv}.  The plane
wave  Born approximation (PWBA)  calculation includes  scattering from
the neutron with detection of  the spectator proton. (As our kinematics
involve  relatively  high  momentum  transfers  and  are  centered  on
$p_m=0$,  the  PWBA  calculation  is  nearly  identical  to  the  PWIA
calculation  which only  includes  scattering from  the proton.)   The
distorted  wave Born  approximation (DWBA)  includes $pn$  final-state
rescattering  (FSI).    The  DWBA+MEC+IC  calculation   includes  also
non-nucleonic currents  (MEC and IC) and the  full calculation 
(DWBA+MEC+IC+RC) further
includes relativistic  contributions of leading order in  $p/m$ to the
kinematical wave  function boost and  to the nucleon current.   The Bonn
two-body interaction \cite{bonn} and  
the Lomon GKex(02S) nucleon form factors \cite{lomon} were
used.  The  models were  acceptance averaged using  MCEEP \cite{mceep}
via  interpolation  over  a  kinematical grid.  The polarizations computed by
Arenh\"ovel were rotated from the center-of-mass system
into the coordinate system of Fig.~\ref{fig:kin}
within MCEEP.  Radiative  folding  was
carried out  within the framework of Borie  and Drechsel \cite{borie}.
It can be seen that the predicted nuclear  effects  are quite  
small for  these kinematics.  However, the  full calculation does not 
give the correct $Q^2$ dependence for \pz{}.
The $\chi^2$ per degree of freedom of the three \pz{} data points relative
to the full calculation is 5.9/3, implying a 12\% probability that our
data are consistent with the theory.  Given the somewhat poorer statistical
uncertainties, the $\chi^2$ per degree of freedom for the double ratio 
deviates from the full calculation by 3.9/3, implying a 27\% probability of
consistency.  As can be seen from Fig.~\ref{fig:hrat}, our highest $Q^2$ 
\hevepv{} datum lies above the world average.  Coupled with the relatively
larger uncertainty of this datum, the
double ratio at this $Q^2$ agrees better with theory than 
the single ratio, \polrat{}. 
It should be cautioned that the lowest $Q^2$ point is the only one 
within the
proton kinetic energy range used to determine the Bonn potential.
Two-photon exchange processes, not included in our calculations, are estimated
to have only minor effects on the transferred polarizations in the 
elastic \hevepv{} reaction \cite{Bl05}.  The effects on
\px{}  and \pz{} are  estimated to  be less  than 0.5\%  for $Q^2=1$  over the
entire $\epsilon$ (longitudinal photon polarization) range.
Since our \devepv{} kinematics are on the quasifree peak, we expect the effects
of two-photon exchange to be of similar size.

\begin{figure}
\includegraphics[width=18pc,angle=0]{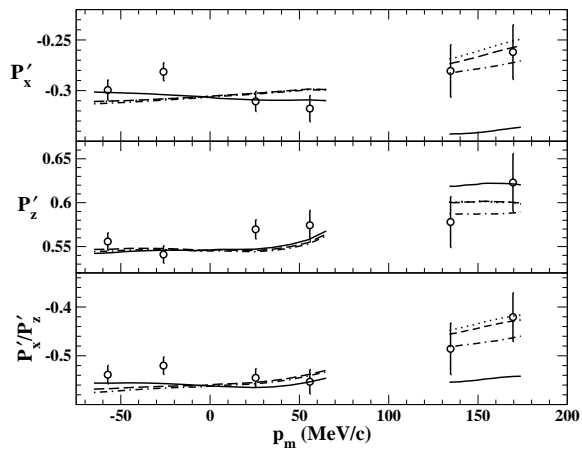}
\caption{\label{fig:ptpl160}  The  polarizations \px{},  \pz{}  and the  ratio
  \polrat{}  as a  function of  $p_m$ at $Q^2=1.00$ \gevsq{}.  
  The $p_m$  values shown correspond to cross section
  weighted  averages.  The  labeling of the theoretical curves
  is the  same as  for the
  previous  figure.    At  low  $p_m$   all  curves  except  for   the  solid
  (DWBA+MEC+IC+RC)  are  essentially  indistinguishable.   For \pz{}  at  high
  $p_m$   the   dotted   (DWBA)    and   dashed   (DWBA+MEC+IC)   curves   are
  indistinguishable.} 
\end{figure}

In  Fig.~\ref{fig:ptpl160}  and  Table~\ref{pr160_tbl}  the  $p_m$ dependence
of the polarizations, \px{} and \pz{}, as well as the polarization ratio,
\polrat{}, is shown for $Q^2=1.00$  \gevsq{}.  Only the
statistical  uncertainties  are plotted  in  the  figure; the  relatively
smaller systematic uncertainties are  given in the table caption.
The group of points at low $p_m$ were obtained by binning the data
for the $p_m=0$ kinematics while the pair of data points
at higher $p_m$ were obtained by binning the data for the 
$p_m=174$ MeV/c kinematics.  The proton spectrometer angles differ
between the two kinematics which gives rise to the discontinuities 
in the calculations between low and high $p_m$.
At low $p_m$ nuclear effects are predicted to have little influence which
is consistent with the results shown in Fig.~\ref{fig:ptpl}.  This is
expected since the latter represents an average over the four low $p_m$ points
in Fig.~\ref{fig:ptpl160}.  At high $p_m$ nuclear effects and especially 
relativistic effects are significantly larger.  
For \pz{} at high $p_m$, the data and full calculation
agree while for \px{} there is a 3.5$\sigma$ discrepancy,
after combining the two highest $p_m$ data points.

The  discrepancy  observed at  our  high  $p_m$  kinematics may  have  serious
implications for the \devenv{} experiment.  In fact, since nuclear effects are
predicted  to  be  larger  for  the  neutron  experiment  (comparison  between
Arenh\"ovel's calculations  for the present  experiment and for  the \devenv{}
experiment \cite{genprc}  suggest that nuclear  effects are four to  six times
larger  for the  neutron  case at  the  lowest and  highest $Q^2$  kinematics,
respectively),  one might  expect any  deviation  from the  calculation to  be
larger as  well.  Without  knowledge of the  dependence of the  discrepancy on
$p_m$  and  on the  out-of-plane  angle  (see  Fig.~\ref{fig:kin}) one  cannot
quantitatively assess the effect on the neutron experiment.  However,
under certain assumptions, one can make an estimate.  To this end, we assume 
that the discrepancy is proportional
to  $p_m$ (and  therefore  zero at  $p_m=0$)  and has  no  dependence on  the
out-of-plane angle.  In this case, our discrepancy would imply a (6$\pm$2)\%
effect on
the neutron form factor at the intermediate $Q^2$, where we have weighted over
the  acceptance of  the neutron  experiment.  This  assumes that  there  is no
magnification in  the effect between the \devepv{}  and \devenv{} experiments.
If, on the other hand, we use the ratio of nuclear effects within the model of
Arenh\"ovel as  a guide, the  effect on the  neutron form factor  increases to
(27$\pm$8)\%.  We caution that these estimates involve a host of 
assumptions.  Only additional data can answer the question definitively. 

\begin{table}
\caption{\label{pr160_tbl} The polarizations \px{}, \pz{} and the ratio
  \polrat{} along with their statistical uncertainties 
as a function of $p_m$ at $Q^2=1.00$ \gevsq{}.  
The statistical uncertainty includes  a contribution from the
statistical uncertainty in $A_c$,
amounting to 1.35\%, except for the highest two $p_m$ points
where it is negligible.
The systematic uncertainties are essentially independent of $p_m$ and are
  estimated to be 0.004, 0.002 and 0.008 for \px{}, \pz{} and \polrat{}
  respectively. }
\begin{ruledtabular}
\begin{tabular}{rccc}
$p_m$ & \px                  & \pz               & \polrat \\
MeV/c &                      &                   &         \\ \hline
$-$57 & $-$0.299 $\pm$ 0.010 & 0.556 $\pm$ 0.013 & $-$0.539 $\pm$ 0.019 \\
$-$26 & $-$0.281 $\pm$ 0.009 & 0.541 $\pm$ 0.013 & $-$0.520 $\pm$ 0.018 \\
   26 & $-$0.311 $\pm$ 0.010 & 0.570 $\pm$ 0.013 & $-$0.545 $\pm$ 0.019 \\
   56 & $-$0.318 $\pm$ 0.013 & 0.574 $\pm$ 0.017 & $-$0.553 $\pm$ 0.025 \\
  135 & $-$0.281 $\pm$ 0.026 & 0.578 $\pm$ 0.029 & $-$0.485 $\pm$ 0.052 \\
  170 & $-$0.262 $\pm$ 0.027 & 0.623 $\pm$ 0.033 & $-$0.420 $\pm$ 0.050 \\
\end{tabular}
\end{ruledtabular}
\end{table}

The breakdown of the systematic uncertainties for \px{}, \pz{}, \polrat{} 
and \doublerat{} is given in Table~\ref{systunc_tbl}.  The uncertainties
are  dominated by uncertainty
in the  precession of the proton's  spin in the  spectrometer magnetic fields.
The spin  precession is characterized by  a rotation matrix  which relates the
polarizations measured with  the FPP to the polarizations  at the experimental
target, \px{} and  \pz{}.  The matrix was obtained  using the COSY \cite{cosy}
transport  program  applied to  the  magnetic elements  of  the  Hall A  Right
spectrometer.  While COSY employs a differential algebraic method to calculate
the transfer matrix, the spin matrix  can also be calculated using a geometric
model~\cite{Pu05}.  In the latter approach the elements of the spin matrix are
based  on   the  proton's  bend   angles  in  the  spectrometer.    Since  the
uncertainties in  the bend  angle can be  measured, this  approach facilitates
estimation of  the precession-related systematic  uncertainties.  So, although
COSY was used  to extract the target polarizations from  those measured at the
FPP,   the  geometric   model  was   employed  to   estimate   our  systematic
uncertainties.  In order to improve the knowledge of 
systematics for the general program
of  Hall~A recoil  polarization  experiments, two  dedicated experiments  were
conducted to  determine the magnitude  of the bend  angle in the
non-dispersive plane along
with  its uncertainty.   The  uncertainty  of the  bend angle in the
dispersive plane was
measured  independently  during   the  experiment  of  Ref.~\cite{Pu05}.   The
geometric  model   was  then  used   to  estimate  the   resulting  systematic
uncertainties on \px{} and \pz{} (the systematic uncertainties on \px{} and
\pz{} are dominated  by uncertainties in the bend  angle in the non-dispersive
and  dispersive planes,  respectively).   For the  double ratio,  \doublerat{},
the systematic  uncertainty  almost completely  cancels since  the
outgoing  protons from  both  reactions travel  through  essentially the  same
magnetic fields.  Finally, especially for the lowest $Q^2$ measurement, 
uncertainty in knowledge of the azimuthal angle of the proton in the FPP
makes a significant contribution to the overall systematic uncertainty.

For \hevepv{}, both \px{} and \pz{} depend on the product $hA_c$ (where $h$ is
the beam  polarization and $A_c$  is the analyzing  power of the FPP)  and the
proton  form   factor  ratio,  $G_E/G_M$.   Therefore,   measurement  of  both
polarization components in \hevepv{} allows determination of $G_E/G_M$ and the
product  $hA_c$.  The  analyzing power  can then  be determined  since  $h$ is
measured  independently   with  the  M{\o}ller  polarimeter.    Note  that  an
uncertainty in  $h$ induces an  uncertainty in $A_c$.  However,  assuming that
$h$  does   not  change  between  the  consecutive   \hevepv{}  and  \devepv{}
measurements, any uncertainty in  this quantity will completely cancel against
the induced uncertainty in $A_c$ in  our extraction of \px{} and \pz{} for the
\devepv{} measurement.   Our extraction  of $A_c$ is  mostly sensitive  to the
uncertainty  in \pz{}  and therefore  to  uncertainty in  the dispersive  bend
angle.   However, an  uncertainty in  the  dispersive bend  angle will  induce
uncertainties in both $A_c$ and \pz{} for \devepv{} which partially cancel one
another, thus,  effectively reducing the  contribution of the  dispersive bend
angle  to  the  total  systematic  uncertainty on  \pz{}.   In  contrast,  the
analyzing  power is  relatively  insensitive  to \px{}  and  therefore to  the
uncertainty  in the  non-dispersive bend  angle  and so  no such  compensation
exists  for \px{}.  Therefore,  the systematic  uncertainty in  \px{} receives
contributions  from  both  $A_c$  and  the  non-dispersive  bend  angle.   The
analyzing  power cancels  in \polrat{}  and so  the systematic  uncertainty on
\polrat{} receives  contributions from both the  dispersive and non-dispersive
bend angles.

\begin{table}
\caption{\label{systunc_tbl} The breakdown of systematic uncertainties
for each kinematics.  The values shown represent absolute uncertainties on the
various quantities.  Here $\theta_{\rm bend}$ and $\phi_{\rm bend}$ refer
to the uncertainties arising from imperfect knowledge of the dispersive
and non-dispersive bend angles in the spectrometer, respectively, while
$\phi_{\rm FPP}$ denotes the uncertainty from the azimuthal angle in the FPP.
The $\theta_{\rm bend}$ contribution to the uncertainty in \px{} is
dominated by the uncertainty in our extraction of the analyzing power (see
the text for details).
The ``Total'' uncertainty is the quadrature sum of the various contributions.
Note that, due to correlations, the uncertainty in \polrat{} is not simply 
the quadrature sum of the uncertainties in \px{} and \pz{}.
}
\begin{ruledtabular}
\begin{tabular}{ccccc}
$Q^2 = 0.43$ (GeV/c)$^2$ &  \px{}  &  \pz{}  & \polrat{} & \fdoublerat{} \\
$p_m = 0$ MeV/c \\
\hline
$\theta_{\rm bend}$      & 0.00000 & 0.00070 & 0.0029    & 0.00005 \\
$\phi_{\rm bend}$        & 0.00015 & 0.00015 & 0.0015    & 0.00045 \\
$\phi_{\rm FPP}$         & 0.00050 & 0.00050 & 0.0037    & 0.00000 \\
Total                    & 0.00056 & 0.00087 & 0.0050    & 0.00045 \\
\\
\hline
$Q^2 = 1.00$ (GeV/c)$^2$ &  \px{}  &  \pz{}  & \polrat{} & \fdoublerat{}
\\
$p_m = 0$ MeV/c \\
\hline
$\theta_{\rm bend}$      & 0.0029  & 0.0027  & 0.0074    & 0.00006 \\
$\phi_{\rm bend}$        & 0.0005  & 0.0003  & 0.0012    & 0.00064 \\
$\phi_{\rm FPP}$         & 0.0009  & 0.0007  & 0.0024    & 0.00018 \\
Total                    & 0.0031  & 0.0028  & 0.0079    & 0.00067 \\
\\
\hline
$Q^2 = 1.61$ (GeV/c)$^2$ &  \px{}  &  \pz{}  & \polrat{} & \fdoublerat{}
\\
$p_m = 0$ MeV/c \\
\hline
$\theta_{\rm bend}$      & 0.011   & 0.0040  & 0.016     & 0.0014 \\
$\phi_{\rm bend}$        & 0.001   & 0.0002  & 0.001     & 0.0006 \\
$\phi_{\rm FPP}$         & 0.001   & 0.0011  & 0.002     & 0.0000 \\
Total                    & 0.011   & 0.0042  & 0.016     & 0.0015 \\
\\
\hline
$Q^2 = 1.00$ (GeV/c)$^2$ &  \px{}  &  \pz{}  & \polrat{} & 
\\
$p_m = 174$ MeV/c \\
\hline
$\theta_{\rm bend}$      & 0.0038  & 0.0019  & 0.0071    &  \\
$\phi_{\rm bend}$        & 0.0007  & 0.0002  & 0.0023    &  \\
$\phi_{\rm FPP}$         & 0.0011  & 0.0007  & 0.0023    &  \\
Total                    & 0.0040  & 0.0020  & 0.0078    &  \\
\end{tabular}
\end{ruledtabular}
\end{table}

In  conclusion, we  have measured  the  \devepv{} and  \hevepv{} reactions  at
$Q^2=0.43$, 1.00 and 1.61 \gevsq{}  for $p_m=0$ and at $Q^2=1.00$ \gevsq{} for
$p_m$ up to 170 MeV/c in Hall A of JLab.  At low $p_m$, the longitudinal 
polarization, \pz{}, exhibits a $Q^2$
dependence at variance with the reaction model for the deuteron.
At high $p_m$, the same model fails to describe the transverse polarization,
\px{}.  These discrepancies indicate that nuclear effects in the \devepv{} 
reaction are not thoroughly understood and further study of this reaction is 
needed.  The discrepancies also suggest
that nuclear corrections in the related neutron electric form factor
experiments need to be studied further.

We acknowledge the outstanding support of the staff of the Accelerator
and  Physics   Divisions  at  Jefferson  Laboratory   that  made  this
experiment  successful.   We also acknowledge useful suggestions of
R. Schiavilla.  This  work  was  supported in  part  by  the
U.S. Department of Energy  Contract No.  DE-AC05-84ER40150 Modification
No. M175 under which
the Southeastern Universities Research Association (SURA) operates the
Thomas  Jefferson National Accelerator  Facility.  We acknowledge additional
grants from the U.S. DOE and NSF, the Italian INFN, the Canadian NSERC
and  the Deutsche Forschungsgemeinschaft (SFB 443).

\end{document}